\documentclass[reprint,twocolumn, 10.5pt, tightenlines,nofootinbib,
notitlepage,longbibliography]{revtex4-1}
\usepackage[left=1in, right=1in,top=1in,bottom=1in,]{geometry}
\usepackage{graphicx} 
\usepackage{float}
\usepackage{array} 
\usepackage{fancyhdr} 
\usepackage{qcircuit}
\usepackage{amsmath,amsthm}
\usepackage{mathtools}
\usepackage{xcolor}

{\bfseries}{\itshape}
\newcommand{\mH}{\mathcal{H}}

\newcommand{\ket}[1]{\ensuremath{\left|#1\right\rangle}} 
\newcommand{\bra}[1]{\ensuremath{\left\langle#1\right|}} 

\renewcommand{\bf}[1]{\ensuremath{\mathbf{#1}}}
\begin{document}
\title{Context aware quantum simulation of a matrix stored in quantum memory} 
\author{Ammar~Daskin}
\affiliation{Department of Computer Engineering, Istanbul Medeniyet University, Uskudar, Istanbul, Turkey}

\author{Teng~Bian}
\author{Rongxin~Xia}
\author{Sabre~Kais}
\affiliation{Department of Chemistry, Department of Physics and Birck Nanotechnology Center, Purdue University, West Lafayette, IN, USA}
\begin{abstract}
In this paper a storage method and a context-aware circuit simulation idea are presented for the sum of block diagonal matrices. Using the design technique for a generalized circuit for the Hamiltonian dynamics through the truncated series, we generalize the idea to (0-1) matrices and discuss the generalization for the real matrices. 
The presented circuit requires $O(n)$ number of quantum gates and yields the correct output with the success probability depending on the number of elements: for matrices with $poly(n)$, the success probability is $1/poly(n)$. 
Since the operations on the circuit are controlled by the data itself, the circuit can be considered as a context aware computing gadget. 
 In addition, it can be used in variational quantum eigensolver and in the simulation of molecular Hamiltonians.
\end{abstract}
\maketitle 
\section{Introduction}
A quantum algorithm  can be described through matrix-vector transformations.  
The number of two-and single-qubit quantum gates required to implement these transformations as a quantum circuit describes the computational complexity of the algorithm.
It is known that an $N\times N$ matrix that depends on $N^2$ independent parameters requires $O(N^2)$ quantum gates \cite{nielsen2010quantum}.
When the matrix is sparse with only $polylog(N)$ nonzero elements, then it is possible to design matrix specific quantum circuits with $polylog(N)$ quantum gates. 
The common circuit design approach is to write the matrix $\mH$ as a sum $\sum_jH_j$, where $e^{iH_jt}$ should be easy to compute for any $H_j$,  and approximate $exp(i\mH t)$ by using the Trotter formula for the exponentiation: i.e. $(\prod_j exp(iH_jt/r)^r$. 
This idea  is used in previous works such as \cite{childs2010simulating, berry2017exponential} to find the most efficient circuits for sparse matrices.

Solutions of many problems relate to finding the lowest (or highest) eigenvalue in magnitude and its associated eigenvector. 
This includes finding the ground state of the Hamiltonian of a quantum system in quantum chemistry \cite{kassal2011simulating,olson2017quantum, kais_book, bian2019quantum, xia2017electronic, xia2018quantum}.
Recent methods such as variational quantum eigensolver \cite{peruzzo2014variational, mcclean2016theory} and quantum signal processing \cite{low2017optimal} have proved that the exponential $e^{i\mH t}$ is not  needed to find the eigenpair of $\mH$. 
For these methods, it is  sufficient to find a direct circuit for $\mH$ that can generate $\mH\ket{\psi}$ for any quantum state \ket{\psi} (example circuits \cite{daskin2012universal,daskin2018direct}).

In this paper, we describe a storage method for (0-1) matrices on quantum memory and show an efficient circuit design that loads the data from quantum memory as a superpositioned state and generates the output $\mH\ket{\psi}$ for any 
 $\mH$  by using quantum operations that are controlled by the data itself. 
When a system uses the context to provide relevant information, it may be considered as a context-aware system \cite{abowd1999towards}.
Therefore, we believe this work will pave the way for quantum context-aware computing.
In terms of the computational complexity, the circuit uses only $O(n)$ number of quantum gates. 
The success probability of the method-as expected-scales with the number of elements and is $1/poly(n)$ for the matrices with $poly(n)$ number of nonzero elements.

This paper is organized as follows: We explain the approach for (0-1) block diagonal matrices in Sec.\ref{Sec2}. In Sec.\ref{Sec3}, we generalize the idea to matrices with (0-1) elements. In Sec.\ref{Sec4}, we analyze the success probability and the storage complexity for sparse matrices with (0-1) elements.  Sec.\ref{Sec5} briefly discusses how the idea generalizes to a matrix in real space, how the circuit can be used with variational quantum eigensolver.

\section{The approach for (0-1) block diagonal matrices}
\label{Sec2}
Assume that we want the circuit implementation of the following  
$N\times N$ matrix:
\begin{equation}
\mH = \sum_i Q_i
\end{equation}
where $Q_i$ is a block diagonal matrix defined as the direct sum of quantum gates: For instance, we will assume $Q_i = \bigoplus_k^K \sigma_k$ with 
$\sigma_k \in \{X, Z, -Z, XZ, ZX, -I, I, Zero\}$, here $Zero$ represents a 2x2 zero matrix and $\bigoplus$ is used for the direct sum of the matrices (It generates a block diagonal matrix.). 
We use a quantum state whose elements encode the gate information of the matrices on the diagonal:
\begin{equation}
\label{Eqgi}
\ket{g_i} = \frac{1}{\eta}\sum_{k=0}^{K-1} \ket{g_{ik}}\ket{k}, 
\end{equation}
where $\eta$ is the normalization constant  and $\ket{g_{ik}}$ is a vector of size 4 from the standard basis.  As an example,
if
\begin{equation}
\label{EqQi}
Q_i = \left(\begin{matrix}
Z&&\\
&X&\\
&&Zero
\end{matrix}\right),
\end{equation} 
then 
\begin{equation}
\label{EqGateEncoding}
\begin{split}
\ket{g_{i0}} = \ket{\bf 1},
\ket{g_{i1}} = \ket{\bf 0}, \\
\ket{g_{i2}} = \left( \begin{matrix}
0\\ \vdots \\0
\end{matrix}\right), 
\text{ and }\eta = \frac{1}{\sqrt{2}}.
\end{split}
\end{equation}
Here, \ket{\bf i} represents the $i$th vector in the standard basis. The zero vector for \ket{g_{i2}} means that it is not included in the summation given in \eqref{Eqgi}.
Notice that \ket{g_i} involves $n+2$ number of qubits: the last $n-1$ qubits indicate the value $k$ and the first three qubits are for the gate-type $g_{ik}$. 

 For an arbitrary given $n$-qubit state \ket{\psi}  and \ket{g_i},
 consider the operation $\ket{\beta}=Q_i\ket{\psi}$.
The circuit in Fig.\ref{FigQi} can be used to produce $1/(\eta\sqrt{8})\ket{\beta}$ on  the amplitudes of the following states (see Appendix \ref{AppendixFig1} for the validation of the circuit):
 \begin{equation}
 \label{EqStates}
 \begin{split}
\frac{1}{\eta\sqrt{8}}\ket{000}\sum_{j = 0}^{N/2-1} 
\bigl(\beta_{2j}\ket{\bf{j}}_{k}\ket{\bf{j}0}_{system} \\
+ \beta_{2j+1}\ket{\bf{j}}_{k} \ket{\bf{j}1}_{system}\bigr), 
\end{split}
 \end{equation}
  where \ket{\bf{j}}  represents the $j$th vector in the basis. The coefficient $1/(\eta\sqrt{8})$ comes from \eqref{Eqgi} and the three Hadamard gates used in the circuit. Therefore, the overall success probability is $||\ket{\beta}/1(\eta\sqrt{8})||^2$.
 
Now let us generalize this to $\sum_i Q_i\ket{\psi}$: First we add one more register to represent \ket{i}, then  we give the following initial state to the circuit: 
\begin{equation}
\ket{g} = \frac{1}{\zeta}\sum_{i,k}\ket{i}\ket{g_{ik}}\ket{k}\ket{\psi}, 
\end{equation}  
where ${\zeta}$ is the normalization constant so that all the nonzero elements in $\ket{i}\ket{g_{ik}}\ket{k}$ are equal to $\frac{1}{\zeta}$.
To get the sum, we apply Hadamard gates to the register representing \ket{i}. The resulting circuit is drawn in Fig.\ref{FigSumQi} (see Appendix \ref{AppendixFig2} for the validation). 
If \ket{g_{ik}}s are stored in the quantum memory, the circuit takes only $O(n)$ time.   This complexity does not change much if we change the size of the basis to add more gates to the set.
In this case, the size of the first register \ket{i} determines the number of Hadamard gates and the success probability. We can define the success probability in this general case as: 
  \begin{equation}
  \label{EqPgeneral}
  P_{success} = 
   \left\lVert\frac{\sum_i Q_i\ket{\psi}}{C\zeta}\right\rVert^2,
  \end{equation}
where $C$ is the coefficient determined by the number of Hadamard gates on the circuit. If the number of Hadamards is close to $n$, then we obtain $\approx \frac{1}{N}$ which is exponentially small in the number of qubits. However, in the sparse case one can expect the number of Hadamards to be much smaller than $n$  as shown in Sec.\ref{Sec4}. 

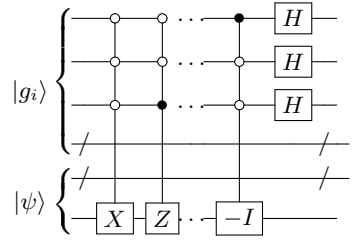
\begin{figure}
	\begin{center}
\mbox{
\Qcircuit @C=.51em @R=.51em {
\lstick{} & \qw &\ctrlo{1} & \ctrlo{1}& \dots& &\ctrl{1} & \gate{H} &\qw & \qw \\
\lstick{} & \qw &\ctrlo{1} & \ctrlo{1}& \dots & &\ctrlo{1} & \gate{H} &\qw & \qw \\
\lstick{} & \qw   & \ctrlo{3}&\ctrl{3}& \dots& &\ctrlo{3}&\gate{H}&\qw  & \qw \\ 
\lstick{} & {/}\qw &\qw	 &\qw&\qw	  & \qw 		&\qw & \qw& {/}\qw & \qw &\ghost{H}
\inputgroupv{1}{4}{0.7em}{3em}{\ket{g_i}}
\\ 
\lstick{} &{/}\qw & \qw	&\qw &\qw&\qw&\qw &\qw &{/}\qw  & \qw & \ghost{H}  \\
\lstick{} & \qw   & \gate{X} & \gate{Z} 	&\dots& & \gate{-I}  &\qw & \qw &\qw 
\inputgroupv{5}{6}{0.7em}{1em}{\ket{\psi}}\\
}
}
\end{center}
\caption{\label{FigQi}Quantum circuit for the implementation of  $Q_i\ket{\psi}$. The circuit includes $2n+2$ qubits and implements the gate set $\{X, Z, -Z, XZ, ZX, -I, I, I\}$.  }
\end{figure}

\begin{figure}
	\begin{center}
		\mbox{
			\Qcircuit @C=.51em @R=.51em {
				\lstick{} &{/} \qw &\qw& \qw&\qw&  \qw&\qw& \gate{\bigotimes H} &{/}\qw & \qw &\rstick{}\\
				\lstick{} & \qw &\ctrlo{1} & \ctrlo{1}& \dots& &\ctrl{1} & \gate{H} &\qw & \qw \\
				\lstick{} & \qw &\ctrlo{1} & \ctrlo{1}& \dots & &\ctrlo{1} & \gate{H} &\qw & \qw \\
				\lstick{} & \qw   & \ctrlo{3}&\ctrl{3}& \dots& &\ctrlo{3}&\gate{H}&\qw  & \qw\\ 
				\lstick{} & {/}\qw &\qw	 &\qw&\qw	  & \qw 		&\qw & \qw& {/}\qw &\qw&\ghost{H} 
				\\
				\lstick{} &{/}\qw & \qw	&\qw &\qw&\qw&\qw &\qw &{/}\qw  & \qw &\ghost{H}  \\
				\lstick{} & \qw   & \gate{X} & \gate{Z} 	&\dots& & \gate{-I}  &\qw & \qw &\qw 
				\inputgroupv{6}{7}{0.7em}{1em}{\ket{\psi}}
				\inputgroupv{1}{5}{0.7em}{4.5em}{\ket{g}}
			}
		}
\end{center}
\caption{\label{FigSumQi}Quantum circuit for the implementation of  $\sum Q_i\ket{\psi}$. 
}
\end{figure}
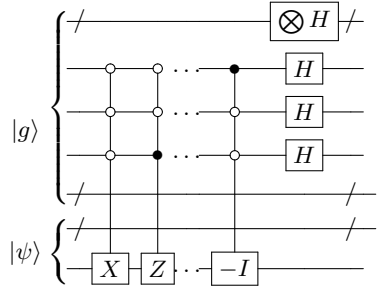

\section{General $\mH$ with 0-1 elements}
\label{Sec3}
In \cite{daskin2018general},  a method is presented to write a general matrix as a sum of unitary matrices. 
In this method, first without changing the location of any element,  two indices $i$ and $k$  are assigned to all $2\times2$ submatrices inside the matrix. For instance, 
\begin{equation}
\label{EqHinAsMatrix}
\mH =	\left( \begin{matrix}
H_{00}& H_{10}& H_{20}&H_{30}\\
H_{11}& H_{01}& H_{31}&H_{21}\\
H_{22}& H_{32}& H_{02}&H_{12}\\
H_{33}& H_{23}& H_{13}&H_{03}
\end{matrix}\right)_{8\times8}.
\end{equation}
Then for $i = 0,\dots,N/2$,  the block diagonal matrix $H_i = \bigoplus_{k=0}^{N/2}H_{ik}$ is constructed. Here, $H_i$ includes one submatrix from each row and $k$ corresponds to the row index of the submatrix (a larger matrix given in \eqref{EqH16x16}, notice the symmetries.).  
$\mH$ is expressed as a sum of $H_i$s in the following form: 
  \begin{equation}
  \mH = \sum_{i=0}^{N/2} H_iP_i. 
  \end{equation}  
Here $P_i$ is a permutation matrix described by using the binary form $i = (b_0...b_{n-1})_2$ as:
\begin{equation}
  P_i =   \left(\bigotimes_{j = 0}^{n-1} X^{b_j}\right) \otimes I.
\end{equation}
 That means an $X$ gate is put on qubit $j$ if there is 1 in the binary representation of $i$. This results in at most $n-1$ single $X$ gates on the circuit.
Here note that while $H_i$ may not be a symmetric matrix, $H_iP_i$ is one.

As in the previous section, we would like to have a circuit where based on the data elements of $H_i$, a control register can choose the set of quantum gates. We will consider sparse matrices  with 0-1 elements and assume that any $H_{ik}$ is in the following form (Note that since the following matrices form a basis, it automatically generalizes to any $H_{ik}$ with 0-1 elements.
):
\begin{equation}
\label{EqGates}
\begin{split}
G_0 =	\left(\begin{matrix}1 & 0\\ 0 &0\end{matrix}\right) 
	&= \frac{I + Z}{2}\\
	G_1=	\left(\begin{matrix}0 & 0\\ 1 &0\end{matrix}\right) 
	&= \frac{X + XZ}{2}\\
	G_2=\left(\begin{matrix}0 & 1\\ 0 &0\end{matrix}\right) 
	&= \frac{X + ZX}{2} \\
G_3 =	\left(\begin{matrix}0 & 0\\ 0 &1\end{matrix}\right) 
	&= \frac{I-Z}{2}\\
Zero = 	\left(\begin{matrix}0 & 0\\ 0 &0\end{matrix}\right) 
	&= Zero+Zero\\
\end{split}
\end{equation}
Using the above assumption,  $H_i$ can be defined as a sum of two block diagonal matrices similarly to $Q_i$ in \eqref{EqQi}: $H_i=\left(Q_{ia}+Q_{ib}\right)$. Then we rewrite $\mH$ as: 
  \begin{equation}
  \mH = \sum_{i=0}^{N/2-1} \left(Q_{ia}P_i+Q_{ib}P_i\right). 
  \end{equation}
For simplicity, we will again assume $Q_{ia}$ and $Q_{ib}$ consist only the gates from the set $\{X,Z,-Z,XZ, ZX,-I,I, Zero\}$ and  use the same encoding as in \eqref{EqGateEncoding}.
Applying $\mH$ to a general quantum state \ket{\psi} leads to a superpositioned state:
 \begin{equation}
 \label{EqHpsi}
\mH\ket{\psi} = \sum_{i=0}^{N/2-1} H_iP_i \ket{\psi}
 \end{equation}
We can construct this superposition state by using a similar circuit to Fig.\ref{FigSumQi}, but in this case we have to control the X gates that implement $P_i$ by the first register representing \ket{i}. The resulting circuit is shown in Fig.\ref{FigHiPipsi}.

 \section{Sparse $\mH$ with $poly(n)$ number of elements}
 \label{Sec4}
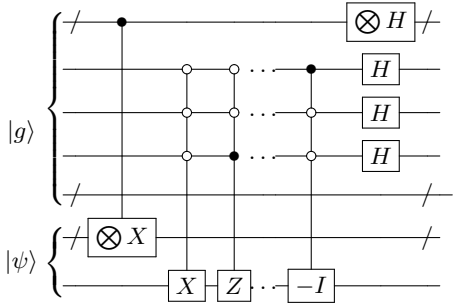
\begin{figure}
	\begin{center}
		\mbox{
			\Qcircuit @C=.51em @R=.51em {
				\lstick{} &{/} \qw & \ctrl{5}&\qw& \qw&\qw&  \qw&\qw& \gate{\bigotimes H} &{/}\qw & \qw &\rstick{}\\
				\lstick{} & \qw &\qw&\ctrlo{1} & \ctrlo{1}& \dots& &\ctrl{1} & \gate{H} &\qw & \qw \\
				\lstick{} & \qw &\qw&\ctrlo{1} & \ctrlo{1}& \dots & &\ctrlo{1} & \gate{H} &\qw & \qw \\
				\lstick{} & \qw   & \qw&\ctrlo{3}&\ctrl{3}& \dots& &\ctrlo{3}&\gate{H}&\qw  & \qw\\ 
				\lstick{} & {/}\qw &\qw&\qw	 &\qw&\qw	  & \qw 		&\qw & \qw& {/}\qw &\qw &\ghost{H}
				\inputgroupv{1}{5}{0.7em}{4.5em}{\ket{g}}
				\\
				\lstick{} &{/}\qw &\gate{\bigotimes X}& \qw	&\qw &\qw&\qw&\qw &\qw &{/}\qw  & \qw   \\
				\lstick{} & \qw   & \qw& \gate{X} & \gate{Z} 	&\dots& & \gate{-I}  &\qw & \qw &\qw 
				\inputgroupv{6}{7}{0.7em}{1em}{\ket{\psi}}\\
			}
		}
\end{center}
\caption{\label{FigHiPipsi}Quantum circuit  implementation of  \eqref{EqHpsi}. The circuit requires $O(n)$ quantum operations. The success probability is given in \eqref{EqPgeneral} for the general case and in \eqref{EqPinSparse} for the sparse case.
}
\end{figure}

For $\mH$ with $poly(n)$ number of 1s, the following two observations can be made:
\begin{enumerate}
\item The number of $H_i$s with only 0-1 elements cannot be more than $poly(n)$. Otherwise, the matrix has more than $poly(n)$ elements or has elements with values different than 0-1s.  Therefore, we load only nonzero $H_i$s and adjust the control bits of the $X$ gates that implements $P_i$s.
This requires only $poly(log(n))$ number of qubits for representing \ket{i} and Hadamard gates.
\item $\zeta$ cannot be more than $poly(n)$. Otherwise, the matrix has more than $poly(n)$ number of nonzero elements. 
\end{enumerate}
Combining these two observations, we can conclude that
  \begin{equation}
  \label{EqPinSparse}
  P_{success} = 
   \left\lVert\frac{\sum_i Q_i\ket{\psi}}{poly(n)}\right\rVert^2.
  \end{equation}
  Therefore, if $\left\lVert\sum_i Q_i\ket{\psi}\right\rVert$ is not too small, the circuit in Fig.\ref{FigHiPipsi} with $O(n)$ number of quantum operations is able to simulate any sparse matrix with $poly(n)$ number of 1s with $O(1/poly(n))$ success probability.
   
\subsection{Storing in quantum memory}
We assume that matrices on the diagonal of $H_i$ are stored as a vector and we can load the superposition of them to the quantum register. That means we store the vectorized form of  $H_{ik}$: 
\begin{equation}
\text{for }H_{ik} = \left(\begin{matrix}0 & 1\\ 0 &0\end{matrix}\right),\   
\ket{g_{ik}}=\left(\begin{matrix}0 \\ 1\\ 0 \\ 0\end{matrix}\right).
\end{equation}
The size of $\ket{g_{ik}}$ depends on the number of quantum gates. Here, it is important to note that if $H_{ik}$ does not have any nonzero element, it is not stored or loaded. 
When $\mH$ has $poly(n)$ number of nonzero elements, most of the $H_i$s must be zero. If $H_i$ does not have any nonzero $H_{ik}$, it is also not stored or loaded.

\section{Discussion}
\label{Sec5}
\subsection{Generalization for  $\mH\in R$}

Suppose we have 
\begin{equation}
\ket{g_{ik}}=\left(\begin{matrix}\alpha_0 \\\alpha_1\\ \alpha_2 \\ \alpha_3\end{matrix}\right),
\end{equation}
where $\alpha$s are the real valued elements of a $H_{ik}$.
The quantum gates $G_0 \dots G_3$s given in \eqref{EqGates} form a computational basis. 
 That means for \ket{g_{ik}} above, it applies the superpositioned of the quantum gates with the probabilities defined by the elements of \ket{g_{ik}}. 
Therefore, after the Hadamard gates on the chosen state, the correct normalized output $\mH\ket{\psi}$  can still be obtained from Fig. \ref{FigHiPipsi} with the probability given in \eqref{EqPgeneral}.
  
\subsection{Use in variational quantum eigensolver }

Variational quantum  eigensolver \cite{peruzzo2014variational, mcclean2016theory} is generally applied to the quantum chemistry problems that are represented by the electronic Hamiltonian in the second quantization by transforming the Hamiltonian to the sum of Pauli operators, which are the products of the Pauli spin matrices(e.g. \cite{o2016scalable,kandala2017hardware,bian2019quantum}). Assume the electronic Hamiltonian is $\mH = \sum_i h_i \mH_i$, where $\mH_i$ is a Pauli operator and $h_i$ is the corresponding coefficient.
The algorithm starts with a state \ket{\psi(\bf{\theta})} defined by the vector of parameters $\bf{\theta}$ and tries to optimize these parameters by minimizing the outcome $\bra{\psi(\bf{\theta})}\mH\ket{\psi(\bf{\theta})} = 
\sum_{i} \bra{\psi(\bf{\theta})} h_i \mH_i \ket{\psi(\bf{\theta})}$.
 Since each $\mH_i$ is assumed to be a product of the Pauli spin matrices, they are implemented by a separate quantum module efficiently.
Therefore, the algorithm involves a quantum part which computes the outcome and a classical part responsible for the optimization by computing the sum of individual outcomes and updating the parameters that forms \ket{\psi}.
The circuit we describe here can be directly used in the quantum part of the algorithm. Since each operator $h_i \mH_i$ can be written as $c_i H_i P_i$, where $c_i$ is $h_i$ or $-h_i$ based on the form of $\mH_i$, $H_i$ is a block diagonal matrix and  $P_i$ is the permutation matrix. In that case, we can simply get:
\begin{equation}
\bra{\psi(\bf{\theta})}\mH\ket{\psi(\bf{\theta})} = 
\sum_{i} \bra{\psi(\bf{\theta})} c_i H_i P_i \ket{\psi(\bf{\theta})}.
 \end{equation} 
This outcome of the circuit can be used in the classical optimization routine to update the parameters of the input state. One can also run each $\bra{\psi(\bf{\theta})} c_i H_i P_i \ket{\psi(\bf{\theta})}$ on separate modules and sum the outcomes in the classical subroutine as done in the original quantum eigensolver.

\begin{figure}[h]
    \centering
    \includegraphics[scale=0.4]{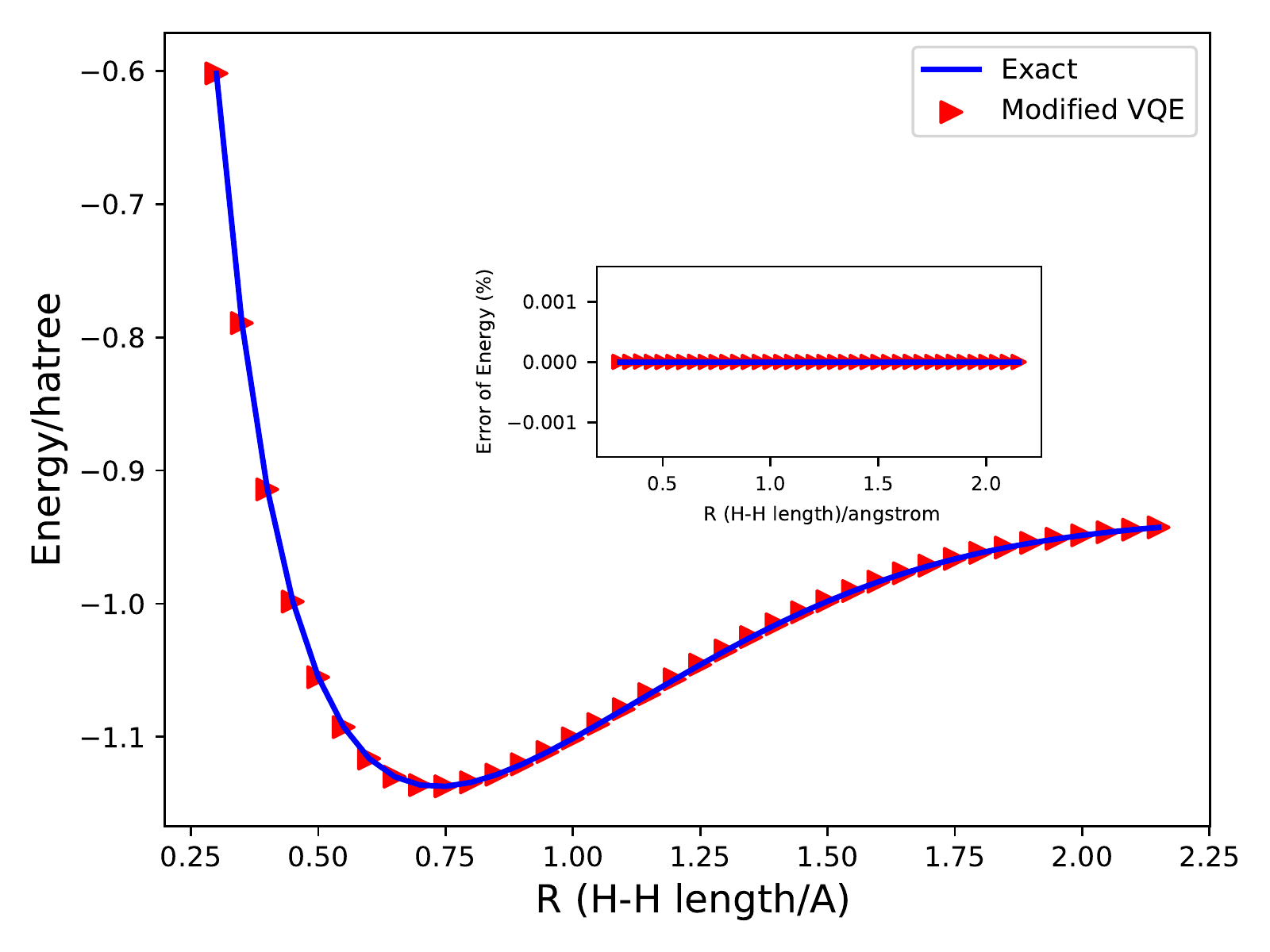}
    \caption{The ground state energy for the H$_2$ molecule as a function of internuclear distance R using the modified VQE based on the contex aware algorithm}
    \label{pairwiseVQE}
\end{figure}

As an example to show how to implement the modified VQE, we will take $h_i \mH_i = h_i XYYZ$. Since $h_i XYYZ$ can be rewritten as $c_i H_i P_i$, in which $P_i = XXXI$, $H_i = IZZZ$ and $c_i = -h_i$. Thus $\bra{\psi(\theta)} h_i XYYZ \ket{\psi(\theta)} = c_i \bra{\psi(\theta)} H_i P_i \ket{\psi(\theta)} $. We can first apply $X$ gates to specific qubits to obtain $P_i\ket{\psi(\theta)}$. Then the context aware algorithm will implement the block diagonal matrix $H_i$, and output the state $H_i P_i \ket{\psi(\theta)}$. By preparing another $\ket{\psi(\theta)}$ and doing quantum fingerprinting\cite{buhrman2001quantum} using swap gate, we obtain $\bra{\psi(\theta)}H_i P_i \ket{\psi(\theta)}$. Summing all the terms with coefficients $c_i$ gives us the energy of state $\ket{\psi(\theta)}$. Then we can use classical optimization methods to update $\theta$ and finally get the minimum energy of the system. FiG. \ref{pairwiseVQE} shows the ground state energy curve of H$_2$ obtained by the simulation based on this modified VQE. In the simulation, the 4-qubit Hamiltonian of H$_2$ is calculated by openfermion package\cite{mcclean2017openfermion} using STO-3G basis set, and the hardware-efficient ansatz is prepared by 3-layer pairwise design in \cite{bian2019quantum}.

\section{Conclusion}
In this paper, we have described a general storage method for matrices and shown circuits that execute quantum gates based on the data information. Since the gates are controlled by the data itself, this idea can be used for context aware quantum computing.
We have also discussed how the circuit can be used with variational quantum eigensolver for any matrix related problems and for the simulation of molecular Hamiltonians. 
As an example, we have shown how the method can be used to calculate the ground state energy curve for the molecule H$_2$, which is in a complete agreement with the exact diagonalization results.
Since any arbitrary real matrix can be decomposed into block diagonal matrices, this method may provide a new way to evolve quantum state by unitary/non-unitary operators through quantum circuits.

\section{Acknowledgement}
One of us, S.K would like to acknowledge the partial support from Purdue Integrative Data Science Initiative and the U.S. Department of Energy, Office of Basic Energy Sciences, under Award Number DE-SC0019215.

\appendix{\section{Validation of the circuit in Fig.\ref{FigQi}}
\label{AppendixFig1}
The controlled gate set in the circuit has the following matrix form:
\begin{equation}
\small    \left(\begin{matrix}
 I^{\otimes 2n-2} \otimes X &&&\\
& I^{\otimes 2n-2} \otimes Z &&\\
&& \ddots &\\
&&& I^{\otimes 2n-2} \otimes I 
\end{matrix}\right)\end{equation}
We can represent the whole circuit more concisely by using the direct sum of matrices:
\begin{equation}
\small
\left( H^{\otimes 3}\otimes I^{\otimes 2n-1}\right)
\left(
\bigoplus_{j=0}^{N/2}X \oplus
\bigoplus_{j=0}^{N/2}Z \oplus
\bigoplus_{j=0}^{N/2}-Z \oplus \dots
\bigoplus_{j=0}^{N/2}I\right)
\end{equation}

To illustrate the action of the circuit, we will use the example matrix $Q_i = Z\oplus XZ$ which leads to the following $\ket{g_{ik}}\ket{k}$s: 
\begin{equation}
\ket{g_{i0}}\ket{0} = \ket{\bf 1}\ket{0}\text{ and } 
\ket{g_{i1}}\ket{1} = \ket{\bf 3}\ket{1}. 
\end{equation}
Then we form the following 6-qubit initial state:
\begin{equation}
\ket{g_i}\ket{\psi} = \frac{1}{\sqrt{2}} \left(\ket{\bf 1}\ket{0}\ket{\psi} + \ket{\bf 3}\ket{1}\ket{\psi}\right).
\end{equation}
After applying the controlled gates ($CG$) to the initial state, we obtain:
\begin{equation}
\small
CG \ket{g_i}\ket{\psi} = \frac{1}{\sqrt{2}} \left(\ket{\bf 1}\ket{0}(Z\otimes Z)\ket{\psi} + \ket{\bf 3}\ket{1}(XZ\otimes XZ)\ket{\psi}\right).
\end{equation}
Applying the Hadamard gates to the first three qubits produces the following final state:
\begin{equation}
\label{EqFig1Out}
\small
\begin{split}
 \frac{1}{4} \bigg(\big(
 \ket{000} - \ket{001}
 +\ket{010}-\ket{011}
 +\ket{100} - \ket{101}\\
 +\ket{110}-\ket{111} \big)
 \ket{0}(Z\otimes Z)\ket{\psi}\\
  + \big( \ket{000} - \ket{001}
 -\ket{010}+\ket{011}
 +\ket{100} - \ket{101}\\
 -\ket{110}+\ket{111} \big)\ket{1}(XZ\otimes XZ)\ket{\psi}\bigg).
\end{split}
\end{equation}
Here, the states where the first three qubits are in \ket{000} includes the expected output which are:
\begin{equation}
\label{EqFig1Out}
\small
 \frac{1}{4} \big(
 \ket{000}\ket{0}(Z\otimes Z)\ket{\psi} 
   +\ket{000}\ket{1}(XZ\otimes XZ)\ket{\psi}  \big).
\end{equation}

The equivalent of $Q_i\ket{\psi}$ is produced on the amplitudes of the states:
\begin{equation}
\label{EqExmChosenStates}
\{\ket{000000}, \ket{000001}, \ket{000110}, \ket{000111}\}.
\end{equation} 

\section{Validations of the circuits in Fig.\ref{FigSumQi} and Fig.\ref{FigHiPipsi}}
\label{AppendixFig2}
In the circuit in Fig.\ref{FigSumQi}, we have the superpositioned input state $\ket{g}$. Before the Hadamard gates on the first register of the circuit,  for different \ket{i} on the output we have  normalized $ Q_i \ket{\psi}$ on the same states as given in \eqref{EqFig1Out} and \eqref{EqExmChosenStates}. That means for \ket{\bf0} on the first register we have normalized $ Q_0 \ket{\psi}$ on the chosen states, and for \ket{\bf1} we have $ Q_1 \ket{\psi}$, and so on.
By applying the Hadamard gates to the first register, for $\ket{i} = \ket{\bf 0}$, we generate the normalized summation $ \sum_i Q_i \ket{\psi}$ on the same chosen states.

Fig.\ref{FigHiPipsi} is just the generalization of Fig.\ref{FigSumQi} and acts the same way.
}
\section{A larger Hamiltonian divided into submatrices}

\begin{equation}
\label{EqH16x16}
\mH =	\left( \begin{matrix}
H_{00}& H_{10}& H_{20}&H_{30}&H_{40}& H_{50}& H_{60}&H_{70}\\
H_{11}& H_{01}& H_{31}&H_{21}&H_{51}& H_{41}& H_{71}&H_{61}\\
H_{22}& H_{32}& H_{02}&H_{12}&H_{62}& H_{72}& H_{42}&H_{52}\\
H_{33}& H_{23}& H_{13}&H_{03}&H_{73}& H_{63}& H_{53}&H_{43}\\
H_{44}& H_{54}& H_{64}&H_{70}&H_{04}& H_{14}& H_{24}&H_{34}\\
H_{55}& H_{45}& H_{75}&H_{61}&H_{15}& H_{05}& H_{35}&H_{25}\\
H_{66}& H_{76}& H_{46}&H_{52}&H_{26}& H_{36}& H_{06}&H_{16}\\
H_{77}& H_{67}& H_{57}&H_{43}&H_{37}& H_{27}& H_{17}&H_{07}\\
\end{matrix}\right)_{16\times16}.
\end{equation}

\bibliographystyle{unsrt}
\bibliography{document2}
\end{document}